\documentclass[12pt, cite, epsf, epsfig]{article}
\usepackage{epsfig}

\fontsize{11}{14}

\title{ Critical Integrated Raman Scattering Intensity near the
cubic-tetragonal phase transition in Strontium Titanate}
\author{Jyoti Dhar Sharma\thanks{Paper presented at the Symposium on 
Neutrons and X-Rays as probes of Condensed Matter: 
A celebration of the work of Professor Roger Cowley, FRS,
(June 30th - July 1st, 2006) at the Department of Physics,
University of Oxford, UK.}\\
Rajiv Gandhi Govt. Degree College, Chaura Maidan, \\ Shimla - 171004,
 HP, India.\\ E-mail: sharmajyotid@yahoo.com}

\begin{document}

\maketitle
\begin{abstract}
Emphasizing the contribution of Professor Roger A Cowley, FRS
to the Theory of  Raman Scattering from crystals,
the development of the Theory of Raman Scattering since 1928
 has been briefly discussed.
 Some experimental studies of Strontium Titanate
  using  Inelastic Neutron Scattering, Raman Scattering, Electro-paramagnetic
   resonance measurement and X-ray \& Gamma Ray techniques has been briefly
   discussed. \\    

Using Schwabl's semi-phenomenological theory for the soft mode and central
 peak,
 we have developed  (a) a  one-phonon Green's function exhibiting the
  three peaked  structure and 
(b) a two phonon Green's function involving one hard mode under damped
quasiharmonic phonon and one three peaked soft-mode phonon. \\

 We have developed the pre-cursor order induced Raman scattering near the
 displacive  phase transition in terms of  Green's functions. \\

Using Group Theory, we have predicted the Raman-active modes in Strontium
Titanate
contributing to Critical Raman Scattering near hard-mode frequencies above
and below
 critical temperature. \\

We have calculated the Critical Integrated Raman Scattering Intensity and the
 Two-phonon Background Raman Scattering Intensity near hard-mode frequencies
  above and below the critical temperature. \\

The results show the same trends as observed in some of the experimental
 observations.
\end{abstract}

\section{The Historical Perspective }

The nature of the spectrum of light scattering from a crystal has been
understood
qualitatively ever since the discovery of the effect in 1928 by C. V. Raman.
Raman scattering  has been extensively studied for several years now
and a vast quantity of both experimental and
theoretical knowledge has been accumulated
\cite{Anderson71, Baker72, Hayes78, Agranovich82, Agranovich84, Cottam89,
Cardona91, Nkoma99}.
Here, we discuss briefly  only the development of the theory of Raman
Scattering from crystals near displacive phase transition
using Green's functions.

 Semiclassical theory of Raman scattering has been discussed by Born and
 Huang \cite{Born56, Sharma81}
 The initial development of the theory of Raman Scattering has been
 reviewed by many authors. \cite{Loudon64, Cowley64, BrucePhD}
An attempt to develop the theory was made by Loudon \cite{Loudon63}.
 He assumed that the Raman scattering by phonons in a crystal occurs through
 a process in which the incident light excites a virtual electron-hole pair, 
which either creates or absorbs one phonon (or more) and then recombines, 
emitting the scattered photon.
Using third order time dependent perturbation theory, he developed expressions for 
the polarizability tensor of the crystal. Based on this approach, Loudon made estimates 
of the Raman Cross section for diamond and zinc sulphide structures which were in 
order of magnitude agreement with the experimental observation. But detailed 
evaluation of the expressions requires knowledge of the electron-band structure 
and electron phonon coupling in the crystal which is not normally available.
Another approach is the expansion technique of the Born and Bradburn \cite{Born47}.
 In this approach it is assumed that the elements of the polarizability tensor can
 be expanded formally in the powers of the displacements of the atoms from their mean 
positions. This technique necessarily involves too many arbitrary parameters and assumptions.

Another attempt to develop the theory of Raman Scattering by phonons was by 
{\bf Cowley in 1964} \cite{Cowley64} and later by { \bf Bruce  and Cowley}
 \cite{Bruce71, Bruce72, BrucePhD}. They developed the theory in terms of
shell model concepts and used it to calculate the two phonon spectrum of alkali halides 
by assuming that the Raman scattering arises from non-linearity in the shell-core
 intra-ion interactions. Quite reasonable agreements with the experimental results 
in NaI and KBr were obtained. The results are considerably improved by using the breathing 
shell model \cite{Schroder66, Nusslein67}.

It has been observed that inter-ionic non-linearities are far less important
for  describing Raman scattering than intra-ionic terms.
Using this model, the two phonon Raman spectrum of SrTi$O_{3}$ in X(YY)Z configuration
 has been calculated by Stirling \cite{Stirling72}. The results give the principal
 features of the experimental spectrum.
Migoni et al.\cite{Migoni76} have used a non-linear shell model to explain the 
second order Raman
 Scattering and temperature dependence of the ferroelectric soft-mode in the AB$O_{3}$
 perovskites. They assume an anisotropy in the Oxygen polarizability. This implies that
 the strong Raman scattering and the behaviour of the ferro-electric soft-mode in the
 perovskites containing Oxygen originate from the universal anisotropic polarizability of 
oxygen. The calculations have been done for SrTi$O_{3}$ and KTa$O_{3}$ and fairly reasonable 
agreement has been obtained with the experimental results for $A_g$ and $E_g$ spectra. 
But the $T_{2g}$ component of the Raman tensor vanishes identically in this model.

The method of linear response theory has also been applied for calculating
the Raman Scatteirng cross sections \cite{Baker72}.

The theory of precursor order induced Raman scattering near displacive phase 
transitions has been developed by Bruce et al. \cite{Bruce80}.

Using Schwabl's theory of soft mode and central peak
\cite{Schwabl72, Schwabl74}, we have  developed
the theory of Raman Scattering near  displacive phase transitions in terms of
Green's Functions and applied it to calculate the Critical Integrated Raman
Scattering Intensity from SrTi$O_3$ near the cubic-tetragonal phase
transition \cite{Sharma81}.
Our results are in close agreement with the experimental observations.

\section{Theory of Raman Scattering near Displacive Phase Transitions}

The Raman scattering of crystal is the inelastic scattering of light by the lattice 
vibrations of the crystal. The Raman tensor is given by \cite{Born56, Sharma81}
\begin{equation}
I_{\alpha \beta \gamma \delta}( \Omega )     = \sum _{ \nu '}< \nu | P_{\alpha \beta}^{*} |
\nu '> < \nu ' | P_{\alpha \beta} | \nu >  \delta ( \Omega - \frac{\xi_{ \nu } - \xi _{ \nu '}}
{ \hbar } )
\end{equation}
where the letters have their usual meanings.
The thermal average is taken over all the initial states $\nu$ .
Using the Heisenberg representation for the operators and an integral representaton for the 
delta function, we get
\begin{equation}
I_{\alpha \beta \gamma \delta}( \Omega ) = \frac{1}{2 \pi N}
\int_{- \infty}^{ \infty}
<  P_{\alpha \beta}^{*}(t) P_{\alpha \beta}(0) >  e^{- i ~ \Omega t}dt
\end{equation}
where N is the number of unit cells in the crystal.

Since the Raman scattering arises from the change in the number of phonons in the crystal,

 we may expand the polarizability operators in the powers of phonon coordinates \cite{Born56}.
For a crystal which undergoes a structural phase transition, we write two expansions.
 We assume that the structural phase ransition takes place when the n-fold degenerate mode 
softens.

1. In the high-temperature phase
\begin{eqnarray}
P_{\alpha \beta}& =& \sum_{qj} P_{\alpha \beta} \left( \matrix
{ q \cr j} \right) Q(qj) \Delta(q) \nonumber \\
& &+ \sum_{q_{1}j_{1} \atop q_{2}j_{2}} P_{\alpha \beta}\left(
\matrix{q_{1} & q_{2}\cr j_{1} & j_{2}} \right) Q(q_{1}j_{1})
Q(q_{2}j_{2})\Delta (q_{1} + q_{2}) + \cdots
\end{eqnarray}

2. In the low-temperature phase

\begin{equation}
\overline{P}_{\alpha \beta} = \sum_{qj} \overline{P}_{\alpha \beta} \left(
\matrix { q \cr j} \right) \overline{Q}(qj)\overline{ \Delta }(q)  + \cdots
\end{equation}

$ Q(qj)$ and $\overline{Q}(qj)$ are the Fourier co-ordinates describing the 
displacements of the atoms from their equilibrium positions above and below 
  the critical temperature $T_{c}$. The presence of the delta functions
indicates  the wave vector conservation.
Some of the modes,which are not Raman-active in the high temperature phase,
 become Raman active below the low temperature phase. In order demonstrate
 this fact, Bruce et al. \cite{Bruce80} derived a relation between 
$ P_{\alpha \beta} \left( \matrix { q \cr j} \right)$ and 
$ \overline{P}_{\alpha \beta} \left( \matrix { q \cr j} \right)$ written as 

\begin{equation}
\overline{P}_{\alpha \beta} \left( \matrix { q \cr j} \right)
\overline{ \Delta }(q) =  P_{\alpha \beta} \left( \matrix { q \cr j}
\right) \Delta(q)+ 2\sum_{j_{s}}Q_{j_{s}} P_{\alpha \beta}\left(
\matrix{q_{s} & -q_{s} \cr j & j_{s}} \right) \Delta (q - q_{s})
\end{equation}

This relation implies that a mode  $(qj)$, for which
$P_{\alpha \beta} \left( \matrix { q \cr j} \right) = 0 $
for $ T > T_{c}$,  becomes raman active in the low temperature
phase if $ q = q_{s}$ and at least one of the coefficients
$P_{\alpha \beta}\left( \matrix{q_{s} & -q_{s} \cr j & j_{s}} \right)$
is non-zero.
Following Bruce et al. \cite{Bruce80}, Raman Tensor describing
the two-phonon Raman scattering (near hard-mode frequencies)
involving the soft mode $(q_{s}j_{s})$ in the high temperature phase is

\begin{eqnarray}
I_{\alpha \beta \gamma \delta}(\Omega) \approx   4\sum_{j_{h}j_{s}}
P_{\alpha \beta}\left( \matrix{q_{s} & -q_{s}\cr j_{h} & j_{s}} \right)
 P_{ \gamma \delta}\left( \matrix{q_{s} & -q_{s}\cr j_{h} & j_{s}} \right)
\sum_{q}\frac{1}{2 \pi N}  \nonumber \\ 
 \int_{- \infty}^{ \infty}
<Q(qj_{s}, 0) Q(-qj_{s}, t)> <Q(qj_{h}, 0) Q(-qj_{h}, t)>
e^{ i ~ \Omega t}dt
\end{eqnarray}

For $ T < T_{c}$, the soft-mode correlation function may be written as

\begin{equation}
<Q(qj_{s}, 0) Q(-qj_{s}, t)> = Q_{j_{s}}^{2} \Delta(q - q_{s})
+ <\overline{Q}(qj_{s}, 0) \overline{Q}(-qj_{s}, t)>
\end{equation}

Defining $A(qj, t)$ as

\begin{equation}
Q(qj, t) = \left( \frac{\hbar }{2\omega(qj)} \right)^{\frac{1}{2}} A(qj,t)
\end{equation}

We get    \\

1. In the high temperature phase

\begin{eqnarray}
I_{\alpha \beta \gamma \delta}( \Omega ) & \approx & 4  \sum_{j_{h}j_{s}}
P_{\alpha \beta}\left( \matrix{q_{s} & -q_{s}\cr j_{h} & j_{s}} \right)
P_{ \gamma \delta}\left( \matrix{q_{s} & -q_{s}\cr j_{h} & j_{s}}
\right) \nonumber \\
& & \sum_{q} \frac{1}{2 \pi N} \frac {\hbar }{2\omega(qj_{s})}
\frac {\hbar }{2\omega(qj_{h})} \nonumber \\ 
 & & \int <A(qj_{s}, 0) A(-qj_{s}, t)> <A(qj_{h}, 0) A(-qj_{h}, t)>
e^{ i ~ \Omega t}dt
\end{eqnarray}
\newpage

2. In the low-temperature phase

\begin{eqnarray}
I_{\alpha \beta \gamma \delta}( \Omega )& \approx&  4 \sum_{j_{h}j_{s}}
P_{\alpha \beta}\left( \matrix{q_{s} & -q_{s}\cr j_{h} & j_{s}} \right)
P_{ \gamma \delta}\left( \matrix{q_{s} & -q_{s}\cr j_{h} & j_{s}} \right)
\sum_{q}\left( \frac{1}{2 \pi N}\right) \left( \frac{\hbar }{2\omega(qj_{s}}
\right)  \nonumber \\
 & &\left( Q_{j_{s}}^{2} \Delta(q - q_{s}) \right.   
\int <A(qj_{h}, 0) A(-qj_{h}, t)> e^{ i ~ \Omega t}dt   
+ \frac {\hbar }{2\omega(qj_{s})}  \nonumber \\ 
 & &\int <A(qj_{s}, 0) A(-qj_{s}, t)> <A(qj_{h}, 0) A(-qj_{s}, t)>
e^{ i ~ \Omega t}dt  \left. \right)
\end{eqnarray}

Eq.(9) describes the two-phonon Raman scattering in the high
temperature phase involving the hard-modes $(qj_{h})$ which couple with
the soft-mode $(qj_{s})$ whereas eq.(10) describes the equivalent one-phonon
and two-phonon Raman Scattering in the low temperature phase. The first term
of eq.(10) describes the one-phonon Raman scattering due to hard-modes which

become Raman active in the low temperature phase.

\section{Green's Functions for three peaked Soft-mode}

Some properties of a crystal may be expressed in terms of
thermodynamic Green's functions
\cite{Abrikosov63, Maradudin62, Cowley63, Cowley66}.
A suitable approximation for the one-phonon Green's function is given
 by {\bf  Cowley} \cite{Cowley66}

\begin{equation}
G(qjj, \Omega + i \epsilon) = \frac{2\omega(qj)}{\beta \hbar}
\left( \frac{1}{\overline{\omega}(qj)^{2} - \omega^{2}
 - 2i\omega(qj)\gamma_{1}} \right)
 \end{equation}
 In terms of Green's functuions, the spectral response function is given by
 \cite{Cowley63}

 \begin{equation}
 \rho (O_{1}O_{2}, \Omega) =   \frac{\beta \hbar}{\pi}
 (1-e^{-\beta \hbar \Omega})ImG(O_{1}O{2}, -\Omega - i \epsilon)
 \end{equation}
 where $O_{1}~and ~o_{2}$ are appropriate operators.
 The one-phonon dynamical form factor is proportional to

 \begin{equation}
 S(qj, \Omega) \propto   \frac{\beta \hbar}{\pi}\frac{1}{2\omega(qj)}
 (1-e^{-\beta \hbar \Omega})ImG(O_{1}O{2}, -\Omega - i \epsilon)
 \end{equation}

 It can be proved that in the limit $\gamma_{1} \longrightarrow 0$,

 \begin{equation}
 Im[G(-\Omega - i \epsilon)] = \frac{1}{\beta \hbar}
 \frac{\omega(qj)}{\overline{\omega}(qj)}[\delta (\Omega - \omega(qj))
 -\delta(\Omega + \omega(qj))]
 \end{equation}
 Hence, we find that 

\begin{eqnarray}
\int <A(qj_{n}, 0) A(-qj_{0}, t)> e^{ i ~ \Omega t}dt & = & \frac{1}{\pi}
 \frac{\omega(qj)}{\overline{\omega}(qj)}( 1 - e^{- \beta \hbar \Omega}) \nonumber \\  & &
  [\delta (\Omega - \omega(qj_{n}))      -\delta(\Omega + \omega(qj_{n}))]
 \end{eqnarray}

 Where A(qj,t) is the  normal mode co-ordinate in the Heisenberg
 representation.
 
\subsection{Schwabl's Theory for Soft-mode and Central-peak}

In 1969, Tani \cite{Tani69} suggested that the soft-mode may develop a
three-peaked structure. Later on { \bf Cowley} \cite{Cowley70}
predicted the presence of a
threepeaked function in the spectral response of a piezoelectric crystal.
According to Cowley, the spectral response of a soft-mode is that of a damped
simple harmonic oscillator coupled to some unknown internal degree of freedom.
\cite{Bruce80}.

A semi-phenomenological theory for the soft-mode and central peak has been
developed by Schwabl \cite{Schwabl72, Schwabl74} which is based on Mori's
theory of Brownian motion \cite{Mori65}. Accordign to this theory, the
dynamic response function exhibing the three-peaked structure may be
written as

\begin{eqnarray}
S(q,\Omega)& = &\Omega \left( 1 - e^{- \beta \hbar \Omega} \right) \chi (q) 
 \left[ \frac{\delta_{c}(q)}{\omega_{\infty}(q)}
\frac{\Gamma_{c}(q)}{\Omega^{2} + \Gamma_{c}^{2}(q)}
  + \left( \frac{\omega_{0}(q)}{\omega_{\infty}(q)} \right)^{2} \right. 
  \nonumber \\ & & \left.
\frac{\omega_{\infty}^{2}(q) \Gamma_{s}(q) - \Gamma_{c}(q)(\Omega^{2} -
\omega_{\infty}^{2}(q)) \left( \frac{\delta_{c}(q)}{\omega_{0}(q)}\right)^{2}}
{(\Omega^{2} - \omega_{\infty}^{2}(q))^{2} +
(\Omega \Gamma_{s}(q))^{2}} \right]
\end{eqnarray}
where letters have their meanings \cite{Sharma81}. $S(q, \Omega)$ has
significant value only when the frequency transfer $\Omega$ is nearly equal to
zero or $\omega_{\infty}$. Therefore we may write

\begin{eqnarray}
S(q,\Omega)& =& \Omega [ 1 - e^{- \beta \hbar \Omega}] \chi (q)
\left[ \left( \frac{\delta_{c}(q)}{\omega_{\infty}(q)}\right)^{2}
\frac{\Gamma_{c}}{\Omega^{2} + \Gamma_{c}^{2}} \right.   \nonumber \\  & &
 \left. +  \frac{\omega_{0}(q)}{\omega_{\infty}(q)}
\frac{\omega_{\infty}^{2}(q) \Gamma_{s}}
{(\Omega^{2} - \omega_{\infty}^{2}(q))^{2} +
(\Omega \Gamma_{s})^{2}} \right]
\end{eqnarray}

\subsection{One-phonon Green's Function for three peaked Soft-mode}
>From Schwabl's theory, we find that the Green's function for three peaked
soft-mode involving central peak is given by

\begin{eqnarray}
G(\Omega + i \epsilon) = -\frac{2 \pi}{ \beta \hbar}
\left( \frac{\delta_{c}(q)}{\omega_{0}(q)}\right)^{2}
\frac{\Gamma_{c}(q)}{\omega _{0}(q)}
\left[ \frac{1}{i(\Omega + i\Gamma_{c})}\right] 
-\frac{\pi}{\beta \hbar} \frac{\omega_{0}(q)}
{\sqrt{\omega_{\infty}^{2}(q) -\frac{1}{4}\Gamma_{s}^{2}}} \nonumber \\
 \left[ \frac{1}{\Omega - \sqrt{\omega_{\infty}^{2}(q)
-\frac{1}{4}\Gamma_{s}^{2}} + \frac{i}{2}\Gamma_{s}}
- \frac{1}{\Omega + \sqrt{\omega_{\infty}^{2}(q)
-\frac{1}{4}\Gamma_{s}^{2}} + \frac{i}{2}\Gamma_{s}} \right]
\end{eqnarray}

\subsection{Two-phonon Green's Function}
The Two-phonon Green's function involving one hard-mode under-damped
quasiharmonic phonon and one soft-mode phonon having three peaked structure
is given by \cite{Coombs73}

\begin{equation}
G(i\omega_{n} = \sum_{i\omega_{m}} G_{1}(i\omega_{m})
G_{2}(i\omega_{n} - i\omega_{m})
\end{equation}.

where $G_{1}$ is the hard-mode phonon and $G_{2}$ is the soft-mode phonon
having three-peaked structure. The sum over $i\omega_{n}$ may be evaluated
using the techniques explained in \cite{Cowley66}. After analytic continuation
(i.e. $ i \omega_{n} \longrightarrow \Omega + i \epsilon $) we get

\begin{eqnarray}
G(\Omega + i\epsilon)& = & \frac{2\pi}{(\beta \hbar)^{2}}
\left( \frac{\delta_{c}(q)}{\omega_{\infty}(q)} \right)^{2}.
\frac{1}{\omega_{0}(q)} \nonumber \\
 & &\left[ \frac{(\Omega - \omega_{h}) - i(\Gamma_{c} + \gamma_{1})}
{(\Omega - \omega_{h})^{2} + (\Gamma_{c} + \gamma_{1})^{2}}
- \frac{(\Omega + \omega_{h}) - i(\Gamma_{c} + \gamma_{1})}
{(\Omega + \omega_{h})^{2} + (\Gamma_{c} + \gamma_{1})^{2}} \right]
\nonumber \\
 & & + \frac{\pi}{(\beta \hbar)^{2}}
\frac{\omega_{0}(q)}{(\sqrt{\omega_{\infty}^{2}
-\frac{1}{2}\Gamma_{s}^{2}}) (\omega_{\infty}^{2}
-\frac{1}{4}\Gamma_{s}^{2})} \nonumber \\
 & & \left[ (\sqrt{\omega_{\infty}^{2} -\frac{1}{4}\Gamma_{s}^{2}}
+ \frac{i}{2}\Gamma_{s}) \left( 
\frac{(-\sqrt{\omega_{\infty}^{2} -\frac{1}{2}\Gamma_{s}^{2}}
-\Omega - \omega_{h}) - i(\frac{1}{2}\Gamma_{s} + \gamma_{1})}
{(\Omega + \sqrt{\omega_{\infty}^{2}-\frac{1}{2}\Gamma_{s}^{2}}
+ \omega_{h})^{2} + (\frac{1}{2}\Gamma_{s} + \gamma_{1})^{2}} \right. \right.
\nonumber \\
  & & \left. - \frac{(-\sqrt{\omega_{\infty}^{2} -\frac{1}{2}\Gamma_{s}^{2}}
-\Omega + \omega_{h}) - i(\frac{1}{2}\Gamma_{s} + \gamma_{1})}
{(\Omega + \sqrt{\omega_{\infty}^{2}-\frac{1}{2}\Gamma_{s}^{2}}
+ \omega_{h})^{2} + (\frac{1}{2}\Gamma_{s} + \gamma_{1})^{2}} \right) 
\nonumber \\
 & & \left. + (\sqrt{\omega_{\infty}^{2} -\frac{1}{4}\Gamma_{s}^{2}}
- \frac{i}{2}\Gamma_{s}) \left( 
\frac{(-\sqrt{\omega_{\infty}^{2} -\frac{1}{2}\Gamma_{s}^{2}}
-\Omega - \omega_{h}) - i(\frac{1}{2}\Gamma_{s} + \gamma_{1})}
{(\Omega + \sqrt{\omega_{\infty}^{2}-\frac{1}{2}\Gamma_{s}^{2}}
+ \omega_{h})^{2} + (\frac{1}{2}\Gamma_{s} + \gamma_{1})^{2}} \right. \right.
\nonumber \\
 & & \left. \left.\frac{(-\sqrt{\omega_{\infty}^{2}
  -\frac{1}{2}\Gamma_{s}^{2}}
-\Omega + \omega_{h}) - i(\frac{1}{2}\Gamma_{s} + \gamma_{1})}
{(\Omega + \sqrt{\omega_{\infty}^{2}-\frac{1}{2}\Gamma_{s}^{2}}
+ \omega_{h})^{2} + (\frac{1}{2}\Gamma_{s} + \gamma_{1})^{2}} \right) \right]
\end{eqnarray}

In the limit $(\gamma_{1} + \Gamma_{c}) \longrightarrow 0$ and
$(\frac{1}{2}\Gamma_{s} + \gamma_{1}) \longrightarrow 0$,
the imaginary part of this Green's function is given by

\begin{eqnarray}
Im[G(-\Omega - i\epsilon)] &=& 2\left( \frac{\pi}{\beta \hbar}\right)^{2}
\left( \frac{\delta_{c}(q)}{\omega_{\infty}(q)}\right)^{2}.
\frac{1}{\omega_{0}(q)}[ -\delta(\Omega + \omega_{h}) +
\delta(\Omega + \omega_{h})]  \nonumber \\
 & &  + \left( \frac{\pi}{\beta \hbar}\right)^{2}
\frac{\omega_{0}(q)}{(\omega_{\infty}(q))^{2}}.
[ - \delta(\Omega + \omega_{\infty}(q) + \omega_{h}) \nonumber \\
 & & + \delta(\Omega + \omega_{\infty}(q) - \omega_{h})
 - \delta(\Omega - \omega_{\infty}(q) + \omega_{h}) \nonumber \\
 & & + \delta(\Omega - \omega_{\infty}(q) - \omega_{h})]
\end{eqnarray}

The intensity of the two-phonon Raman Scattering involving the central
peak contribution may be obtained from the first part of the above expression.

Thus the two-phonon spectral function involving one hard-mode phonon and one
three peaked soft-mode phonon is given by

\begin{eqnarray}
\int <A(qj_{s}, 0) A(-qj_{s}, t)> <A(qj_{h}, 0) A(-qj_{h}, t)>
e^{i\Omega t} dt  \nonumber \\
= \frac{\beta \hbar}{\pi}\left(1- e^{-\beta \hbar \Omega}\right)
\left[ 2 \left( \frac{\pi}{\beta \hbar}\right)^{2}
\left( \frac{\delta_{c}(q)}{\omega_{\infty}(q)} \right)^{2}
\frac{1}{\omega_{0}(q)}\right.\nonumber \\ 
\{ - \delta(\Omega + \omega_{h}) + \delta(\Omega - \omega_{h})\}
+ \left( \frac{\pi}{\beta \hbar}\right)^{2}
\frac{\omega_{0}(q)}{(\omega_{\infty}(q))^{2}} \nonumber \\
\left\{ - \delta(\Omega + \omega_{\infty}(q) + \omega_{h}) 
+ \delta(\Omega + \omega_{\infty}(q) - \omega_{h}) \right. \nonumber \\
\left. \left. - \delta(\Omega - \omega_{\infty}(q) + \omega_{h})
+ \delta(\Omega - \omega_{\infty}(q) - \omega_{h}) \right\} \right]
\end{eqnarray}

\section{ The Raman Tensor in terms of Green's Functions}
Thus we find that in the high temperature phase, the Raman Tensor is
given by

\begin{eqnarray}
I_{\alpha \beta \gamma \delta}( \Omega ) &  \approx & 4  \sum_{j_{h}j_{s}}
P_{\alpha \beta}\left( \matrix{q_{s} & -q_{s}\cr j_{h} & j_{s}} \right)
P_{ \gamma \delta}\left( \matrix{q_{s} & -q_{s}\cr j_{h} & j_{s}}
\right) \nonumber \\
 & & \sum_{q} \frac{1}{2 \pi N} \frac {\hbar }{2\omega(qj_{s})}
\frac {\hbar }{2\omega(qj_{h})}\left(1- e^{-\beta \hbar \Omega}\right) \nonumber \\ 
 & & \left[ 2 \left( \frac{\pi}{\beta \hbar}\right)
\left( \frac{\delta_{c}(q)}{\omega_{\infty}(q)} \right)^{2}
\frac{1}{\omega_{0}(q)}\right.\nonumber \\ 
 & & \{ - \delta(\Omega + \omega_{h}) + \delta(\Omega - \omega_{h})\}
+ \left( \frac{\pi}{\beta \hbar}\right)
\frac{\omega_{0}(q)}{(\omega_{\infty}(q))^{2}} \nonumber \\
 & & \left\{ - \delta(\Omega + \omega_{\infty}(q) + \omega_{h}) 
+ \delta(\Omega + \omega_{\infty}(q) - \omega_{h}) \right. \nonumber \\
 & & \left. \left. - \delta(\Omega - \omega_{\infty}(q) + \omega_{h})
+ \delta(\Omega - \omega_{\infty}(q) - \omega_{h}) \right\} \right]
\end{eqnarray}

 \newpage

Similarly, in the low temperature, we obtain

\begin{eqnarray}
I_{\alpha \beta \gamma \delta}( \Omega ) &  \approx & 4  \sum_{j_{h}j_{s}}
P_{\alpha \beta}\left( \matrix{q_{s} & -q_{s}\cr j_{h} & j_{s}} \right)
P_{ \gamma \delta}\left( \matrix{q_{s} & -q_{s}\cr j_{h} & j_{s}}
\right) \nonumber \\
 & & \sum_{q} \frac{1}{2 \pi N}\frac {\hbar }{2\omega(qj_{h})}
 \left[ Q_{j_{s}}^{2} \Delta(q-q_{s})
\frac{1}{\pi}\frac{\omega(qj_{h})}{\overline{\omega}(qj_{h})}
 \left( 1- e^{-\beta \hbar \Omega}\right) \right. \nonumber \\
 & & \{\delta(\Omega - \omega_{h}) - \delta(\Omega + \omega_{h})\} 
 + \frac {\hbar }{2\omega(qj_{s})}
\left(1- e^{-\beta \hbar \Omega}\right) \nonumber \\
 & & \left[  \left( \frac{2\pi}{\beta \hbar}\right)
\left( \frac{\delta_{c}(q)}{\omega_{\infty}(q)} \right)^{2}
\frac{1}{\omega_{0}(q)}\right.\nonumber \\
 & & \{ - \delta(\Omega + \omega_{h}) + \delta(\Omega - \omega_{h})\}
+ \left( \frac{\pi}{\beta \hbar}\right)
\frac{\omega_{0}(q)}{(\omega_{\infty}(q))^{2}} \nonumber \\
 & & \left\{ - \delta(\Omega + \omega_{\infty}(q) + \omega_{h}) 
+ \delta(\Omega + \omega_{\infty}(q) - \omega_{h}) \right. \nonumber \\
 & & \left. \left. \left.- \delta(\Omega - \omega_{\infty}(q) + \omega_{h})
+ \delta(\Omega - \omega_{\infty}(q) - \omega_{h}) \right\} \right] \right]
\end{eqnarray}

\section{Some Experimental Studies of Strontium Titanate}

\subsection{Crystal Structure of Strontium Titanate in the low
temperature phase}

The lattice dynamics of $SrTiO_{3}$ has been of great experimental and
theoretical interest.

Inelastic neutron scattering methods have been
extensively used to measure the phonon dispersion curves and to investigate
the temperature dependence of the soft-mode frequencies
\cite{Cowley64a, Cowley69, StirlingPhD, Stirling72, Shapiro72}. Also, other
techniques like Raman scattering \cite{Neilsen68, Fleury68, Taylor79},
electron-paramagnetic resonance\cite{Muller71}
measurements, X-ray and$\gamma$ - ray scattering
\cite{Darlington75, Darlington76}
have been used to
investigate the nature of structural phase transition in  $SrTiO_{3}$.
It has been established that  $SrTiO_{3}$ undergoes  cubic to tetragonal
phase transition at 110K. The transition temperature may differ slightly
for different crystals. Shapiro et al. \cite{Shapiro72} quote the transition
temperature to be 100K where as Muller and Berlinger \cite{Muller71} have
indicated to be 105.5K.

A mechanism for this phase transition, based on an accidental degeneracy
of the frequencies of two phonon branches in a wide  q-range,
was proposed by Cowley \cite{Cowley64a}. But for this model, no conclusive
experimental evidence has been given. The generally accepted mechanism of this phase
transition was suggested by Unoki and Sakudo \cite{Unoki67} on the basis of
their ESR measurements. They deduced the low temperature phase crystal
structure from the atomic positions of the oxygen ions. The unit cell of this
structure is $ \sqrt{2}a X \sqrt{2}a X 2c$ and the space group is I4/mcm
$(D_{4h}^{18})$ where a and c correspond to tetragonal unit containing
one molecule. According to Unoki and Sakudo, this structure seemed to suggest
an instability in some optical branch at large wave vector.

Motivated by this low temperature phase structure, Fleury, Scott and Worlock
\cite{Fleury68a} proposed an interpretation of this cubic-tetragonal phase
transition. The basic features of their model are :

(i) The frequency of the triply degenarate $\Gamma_{25}$ mode at the R-point
( also called $R_{25}$-mode) of the Brillouin zone tends to zero as
$ T \longrightarrow T_{c}^{+}$.

(ii) In the low temperature phase, the R-point becomes a reciprocal lattice
point, the unit cell is doubled and there are additional excitations
with wave vectors at the zone-centre.

(iii) As the temperature falls below $T_{c}$, the frequencies of the two
new zone-centre phonons( the progenitors of $\Gamma_{25}$ zone boundary
phonon) increase.

This model was based on the following experimental evidence:

(i) Several sharp lines appear in the Raman spectrum of $SrTiO_{3}$
below the transition temperature \cite{Neilsen68, Taylor79}.

(ii) As $ T \longrightarrow T_{c}$ from below, the two zone-corner phonon
frequenciesdefinitely soften \cite{Fleury68a}.

(iii) In the electric field induced Raman scattering experiments, it has
been observed that the two zone-centre phonons interact directly with
the components of the 'ferroelectric mode' \cite{Fleury68, Worlock67}.

Using the inelastic neutron scattering techniques, Shirane and Yamada
 \cite{Shirane69} proved conclusively that this phase transition is caused
 by an instability in the $\Gamma_{25}$ zone-boundary mode.

\subsection{The temperature dependence of the Soft-mode frequencies
of Strontium Titanate}

The temperature dependence of the Soft-mode frequencies
of  $SrTiO_{3}$ has been studied by neutron-scattering as well as Raman
scattering techniques. In the cubic phase, the $R_{25}$ soft-mode has been
studied by Cowley et al. \cite{Cowley69} and Shapiro et al. \cite{Shapiro72}
using inelastic neutron scattering methods. From the linear portions of
the curve it is observed that

\begin{equation}
\omega_{0}^{2}(0) = a_{0} (T - T_{c}), ~~~  a_{0} \cong 13.37cm^{-2}/K
\end{equation}

In the low temperature phase, Cowley \cite{Cowley69} has studied the
temperature dependance of the soft mode using neutron-scattering  technique.
Fleury et al. have studied the temperature dependence of both the components
of the sof-mode using Raman scattering technique. Using temperature-derivative
Raman spectroscopy, Yacoby et al. \cite{Yacoby76}have studied the $\gamma_{1}
(A_{1g})$ component of the soft-mode below $T_{c}$. Their measurements show
that
\begin{equation}
\omega_{0}^{2} = a_{0}(T_{c}-T), ~~~ a_{0} \approx 29.2cm^{-2}/K
\end{equation}

\subsection{Temperature-dependence of the static displacement of Oxygen atoms
in Strontium Titanate in the low temperature phase}

The phase transition in $SrTiO_{3}$ occurs when, in the cubic phase,
alternate static rotations of nearly rigid $TiO_{6}$ octahedra take place
around [100] axes.The temperature dependance of the rotation angle has been
determined by Muller and Berlinger \cite{Muller71} using paramagnetic resonance
of$Fe^{3+}$ ions substituted for $Ti^{4+}$ in $SrTiO_{3}$. From their
measurements, it is found that the temperature dependence of the normal mode
amplitude corresponding to the static displacements of oxygen atoms in the
soft $R_{25}$-mode is given by

\begin{equation}
Q^{2} = Q_{0}^{2}(T_{c}-T)^{\frac{2}{3}}
\end{equation}

where $$ Q_{0}^{2} = 5.175 X 10^{-42}~~ cm^{2}gm(deg)^{-\frac{2}{3}} $$

\subsection{The Central Peak}

In the neutrino-scattering stydy of the soft-mode in $SrTiO_{3}$, Riste et al. 
\cite{Riste71}, for the first time, observed a central component centred around
$\omega = 0$ and $ q = q_{R}$ in addition to the phonon side bands centered around $ \omega = \omega_{\infty}$. Shapiro et al. \cite{Shapiro72} reported Higher resolution neutron scattering studies of side bands and central component in $SrTiO_{3}$ and also observed central component in $KMnF_{3}$.

The central component in $SrTiO_{3}$ first appears at 70K above $T_{c}$ and
exhibits a critical divergence as $ T \longrightarrow T_{c}^{+}$.
The frequency of the side-bands first softens in accordance with the
soft-mode theory but finally saturates a few degrees above $T_{c}$.
The frequency width of the central component is found to be less
than the instrumental resolution of $5X10^{10}$ Hz. In their high
resolution neutron scattering study of central peak, T$\ddot{o}$pler et al.
\cite{Topler77} observed no frequency broadening or wave vector broadening
within the instrumental resolution. This central peak has also been studied
 by X-ray \cite{Darlington75} and $\gamma$-ray \cite{ Darlington76}
 techniques with excellent energy resolution $(3.6 X 10^{6} Hz)$ but no
 frequency broadening has been observed. Also it
 has been established that it does not originate from the crystal surface
 only \cite{Cowley78}.

\subsection{ The Raman Spectrum of Strontium Titanate}

An early attempt to record the Raman spectrum of $SrTiO_{3}$  was made by
Narayanan and Vedam \cite{Nara61}. They used $\lambda $ 4358 of mercury as
an axciter and observed 7 Raman lines, one of which was of low frequency.
In the cubic structure of $SrTiO_{3}$, there could be no first order
Raman scattering. Narayan and Vedam interpreted six of these as a first
order Raman spectrum arising from a small deviation of the cubic
 $SrTiO_{3}$  from its idealized symmetry. Neilsen and Skinner
 \cite{Neilsen68} have recorded the spectrum measured in the X(YY)Z
 configuration at various temperatures. They used an argon-ion laser
 as the exciting source and an in-tandem double grating spectrum  to
 analyse the scattered light. At room temperature the spectrum was
 found to be second order. Below the phase transition at 110K, three
 sharp lines appeared in the spectrum which were interpreted to be
 due to the scattering from local modes rather than from polar tranverse
 optic modes.

Fleury and Worlock \cite{Fleury68} have studied the electric field induced
Raman scattering from the four TO phonons in $SrTiO_{3}$  at temperatures
between 8K and 300K. They observed that the soft-mode frequency at 8K
increases from 10 $cm^{-1}$ at zero field to 45 $cm^{-1}$ at 12 KV/cm.
Migoni et al. \cite{Migoni76} have recorded the three independent components
of the second-order Raman spectra at 300K. An exhaustive study of the
tetragonal phase of $SrTiO_{3}$  by Raman scattering has been reported by
Taylor and Murray \cite{Taylor79}. They observed the spectra from two
samples of $SrTiO_{3}$  in the X(ZZ)Y and X(ZX)Y configuration above and
below the structural phase transition. Argon and Krypton lasers were used
to excite the spectra. The $90^{0}$ scattered light was analysed by a
computer controlled spectrometer. For the soft-mode spectrum, the spectral
resolution was 1.8 $cm^{-1}$ and for the rest of the spectra it was
2.5 $cm^{-1}$. It has been observed that in addition to the high
intensity multi phonon continuum, the spectra shows four first order peaks.
It is believed that the peaks at 142 $cm^{-1}$ and 444 $cm^{-1}$ arise from
$B_{2g}-E_{g}$ pair of the modes and those at 171 $cm^{-1}$ and
543 $cm^{-1}$, which were found to be sample dependent, result from
impurity induced first order Raman scattering.
 \newpage
\section{Group Theory}
In Solid State Physics, Group Theory is applied to make use of the symmetry
properties of a crystals \cite{Heine60, Tinkham64}. In the context of 
theory of Raman Scattering from crystals, Cowley has discussed \cite{Cowley71a}
the application of Group theory for classifying normal modes of vibration of crystals and obtaining the selection Rulesfor one-phonon and two-phonon 
Raman Scattering.

Using Group Theory, it is possible to predict the number and symmetry of Raman 
active modes in a particular phase and to correlate these with equivalent modes
 of different species in the other phases. Lockwood and Torrie
 \cite{Lockwood74}
have tabulated the branching rules for symmetry species of phonons at high 
symmetry points for a perovskite type crystal which undergoes a cubic-tetragonal
and also a tetragonal-orthorhombic phase transition.

We predict  the two-phonon Raman active combinations in the cubic 
phase contributing to the critical Raman scattering above $T_{c}$ near the 
hard-mode frequencies along with equivalent one-phonon Raman active modes and 
two-phonon Raman active combinations in the tetragonal phase in $SrTiO_{3}$
as shown in Table 1.

The R-point becomes the zone-centre point at the cubic-tetragonal phase transition in $SrTiO_{3}$. The irreducible representations correspond 
 to the Sr atom at the origin to be consistent with Cowley \cite{Cowley64}.
The character labeling system of Koster \cite{Koster57} has been followed.

\begin{table}
\caption{Raman-active modes in Strontium Titanate contributing to
Critical Scattering near hard-mode frequencies above and below
critical temperature.}
\begin{tabular}{||ccc||} \hline  \hline 
Hard-mode    & Two-phonon Raman-active    &  Equivalent one-phonon \\
 frequency   & combinations contributing  & Raman active modes \\
$\omega_{h}$ & to the Critical Raman      & and two-phonon Raman active \\
    & scattering near $\omega_{h}$   & combinations in tetragonal phase,  \\
             &  above $T_{c}$              & contributing near \\
             &                        & $\omega_{h}$ below $T_{c}$ \\ \hline
 & &
$\Gamma_{3}, \Gamma_{5}, \Gamma_{3}X \Gamma_{1}(soft)$   \\
13.5 THz & $R_{25}(soft)X R_{15}$  & $\Gamma_{3}X \Gamma_{5}(soft)$   \\ 
 & &     $\Gamma_{5}X \Gamma_{1}(soft),
 \Gamma_{5}X \Gamma_{5}(soft)$ \\ \hline
 & &
$\Gamma_{4}, \Gamma_{1}(soft)X \Gamma_{4}$   \\
 7.8 THz &   $R_{25}(soft)X R_{12}'$  & $\Gamma_{5}(soft)X \Gamma_{4}$   \\ 
 & &     $\Gamma_{5}(soft)X \Gamma_{2}$ \\ \hline
 & &
$\Gamma_{3}, \Gamma_{5}, \Gamma_{3}X \Gamma_{1}(soft)$   \\
 4.36 THz &   $R_{25}(soft)X R_{15}$  & $\Gamma_{3}X \Gamma_{5}(soft)$   \\ 
 & &     $\Gamma_{5}X \Gamma_{1}(soft),
 \Gamma_{5}X \Gamma_{5}(soft)$ \\ \hline  \hline
\end{tabular}
\end{table}

\section{Critical Integrated Raman Scattering Intensity}

In eq.()and eq(), integration is performed over $\Omega$ using
$\delta$-functuins. The dominant contriburtion near $T_{c}$ is due to the
central-peak and the hard-modes which appear in two-phonon Raman active
combinations involving the soft mode above $T_{c}$

For $T > T_{c}$,

\begin{eqnarray}
\int I^{crit}_{\alpha \beta \gamma \delta}(\Omega) & \approx &
4  \sum_{j_{h}j_{s}}
P_{\alpha \beta}\left( \matrix{q_{s} & -q_{s}\cr j_{h} & j_{s}} \right)
P_{ \gamma \delta}\left( \matrix{q_{s} & -q_{s}\cr j_{h} & j_{s}}
\right) \nonumber \\
 & & \left( \frac{\hbar}{2\omega(0j_{h})} \right) [ 2 n(\omega(0j_{h})) + 1]
\frac{1}{2\beta N} \nonumber \\
 & & \sum_{q} \left( \frac{\delta_{c}(q)}{\omega_{\infty}(qj_{s})} \right)^{2}
\frac{1}{\omega_{0}(qj_{s})}
\end{eqnarray}
$\omega_{\infty}(q)$ and $\omega_{0}(q)$ are related by
\begin{equation}
\omega^{2}_{\infty}(q) = \omega_{0}(q)^{2} + \delta_{c}^{2}(q)
\end{equation}
The sum over q is converted to an integral with a finite cut off $\Lambda$.
For the soft-mode, anisotropic dispersion relation given by Stirling
\cite{Stirling72} has been used.
Thus we obtain
\begin{eqnarray}
\int I^{crit}_{\alpha \beta \gamma \delta}(\Omega) & \approx &
4  \sum_{j_{h}j_{s}}
P_{\alpha \beta}\left( \matrix{q_{s} & -q_{s}\cr j_{h} & j_{s}} \right)
P_{ \gamma \delta}\left( \matrix{q_{s} & -q_{s}\cr j_{h} & j_{s}}
\right) \nonumber \\
 & & \left( \frac{\hbar}{2\omega(0j_{h})} \right) [ 2 n(\omega(0j_{h})) + 1]
 \nonumber \\
 & & \frac{1}{2\beta} ( \sqrt{\omega_{0}^{2}(0) + \delta_{c}^{2}}
 - \omega_{0}(0)) . \overline{A}
\end{eqnarray}

Similarly for $T < T_{c}$ we get

\begin{eqnarray}
\int I^{crit}_{\alpha \beta \gamma \delta}(\Omega) & \approx &
4  \sum_{j_{h}j_{s}}
P_{\alpha \beta}\left( \matrix{q_{s} & -q_{s}\cr j_{h} & j_{s}} \right)
P_{ \gamma \delta}\left( \matrix{q_{s} & -q_{s}\cr j_{h} & j_{s}}
\right) \nonumber \\
& &\frac{1}{2\pi} \left( \frac{\hbar}{2\omega(0j_{h})} \right)
\left \{Q_{j_{s}}^{2}\frac{1}{\pi} [ 2 n(\omega(0j_{h})) + 1] \right.
\nonumber \\ 
 & & + \left. \frac{\pi}{\beta }[ 2 n(\omega(0j_{h})) + 1]
( \sqrt{\omega_{0}^{2}(0) + \delta_{c}^{2}}
 - \omega_{0}(0)) . \overline{A} \right\}
\end{eqnarray}

$n(\omega(qj))$ denotes the occupation number for the mode (qj)
\cite{Cowley66}.

\section{Numerical Calculations}

The numerical calculations have been performed for two-phonon background
Raman scattering intensity
(without central peak) and the Critical Integrated Raman Scattering Intensity
for $SrTiO_{3}$ near the cubic-tetragonal phase
transition \cite{Sharma81}.

It has been assumed that $\Gamma_{1}$ soft-mode has similar three-peaked
structure in the low temperature phase as the $R_{25}$ soft-mode. The
anharmonic coupling constant $\delta$ is assumed to be temperature
independent and equal to 0.1325 THz.
The results are shown in Figs 1, 2 \& 3.

\section{Conclusions}
For the first time, an attempt has been made to calculate the two-phonon
background  Raman Scattering Intensity and the Critical Integrated
Raman Scattering Intensity near the cubic-tetragonal phase transition
in $SrTiO_{3}$.

Our results show the same trends as in  the experimental observations of
Riste et. al \cite{Riste71a, Cowley80a} and E. Courtens
\cite{Courtens72, Bruce80a}.

We have been able to evaluate numerically only the diogonal Raman Tensor
$I_{XXXX}$. For evaluating the off-diagonal Raman Tensor the temperature
dependence of $\Gamma_{5}(E_{g})$ soft-mode is required.
Similar calculations may be performed for the other isomorphs of $SrTiO_{3}$ .

\section*{Acknowledgements}
The aouthor is thankful to Prof. S. Dev and Prof. P. K. Ahluwalia of the
Himachal Pradesh University for
encouragement during the progress of writing this paper.
This work is a part of the thesis  submitted for the degree of M. Phil. at the University
of Edinburgh under the supervision of Prof. R. A. Cowley, FRS and
Dr. A. D. Bruce and the author is thankful to them, especially to
Frof. R. A. Cowley
for guidance, helpful discussions and also for suggesting this
research problem. The Studentship of the University of Edinburgh is
also gratefully acknowledged.

\newpage
\begin{figure}{}
\begin{center}
{\epsfig{file=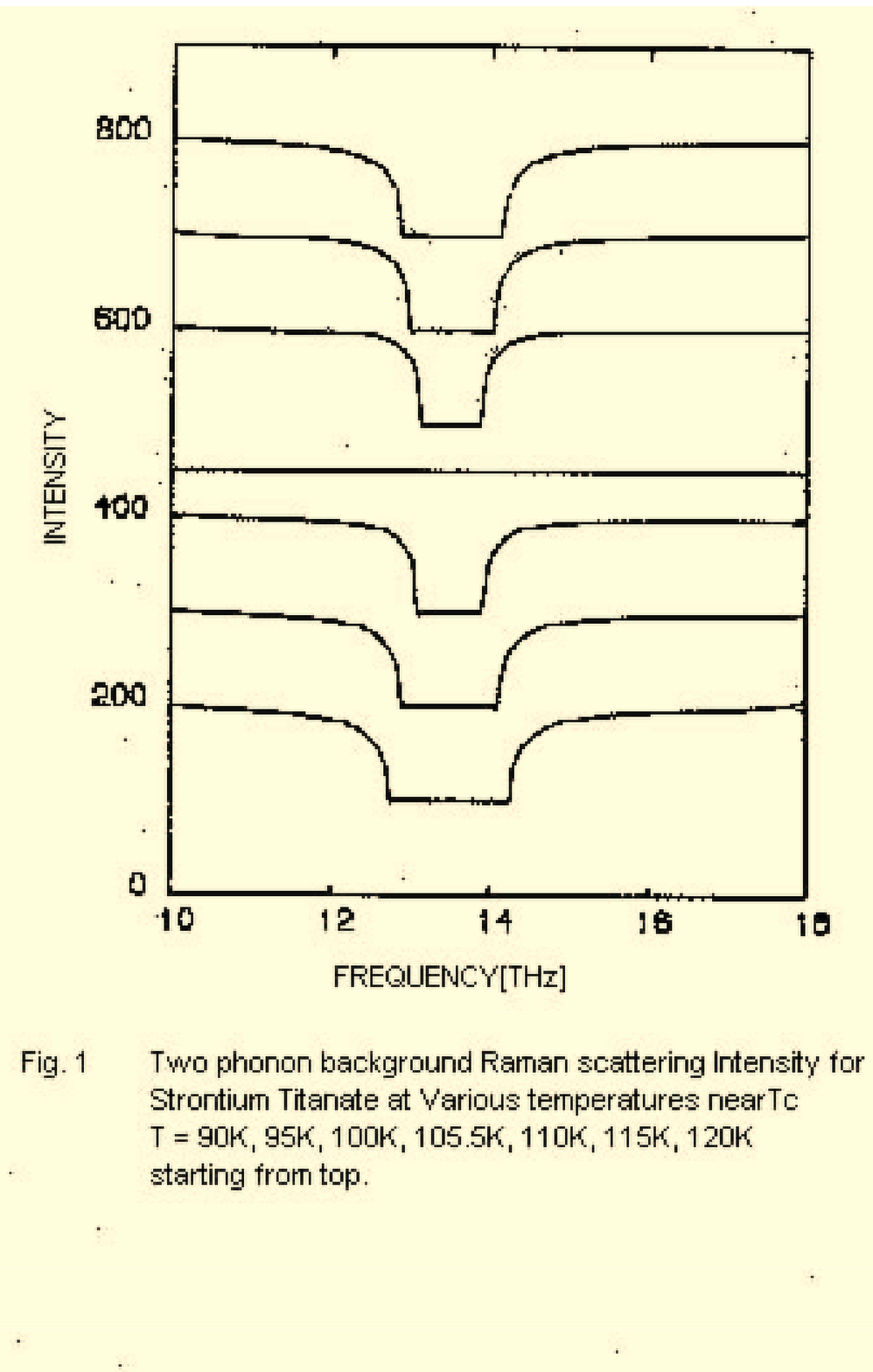, width=8.0cm, height=8.0cm}}
{\epsfig{file=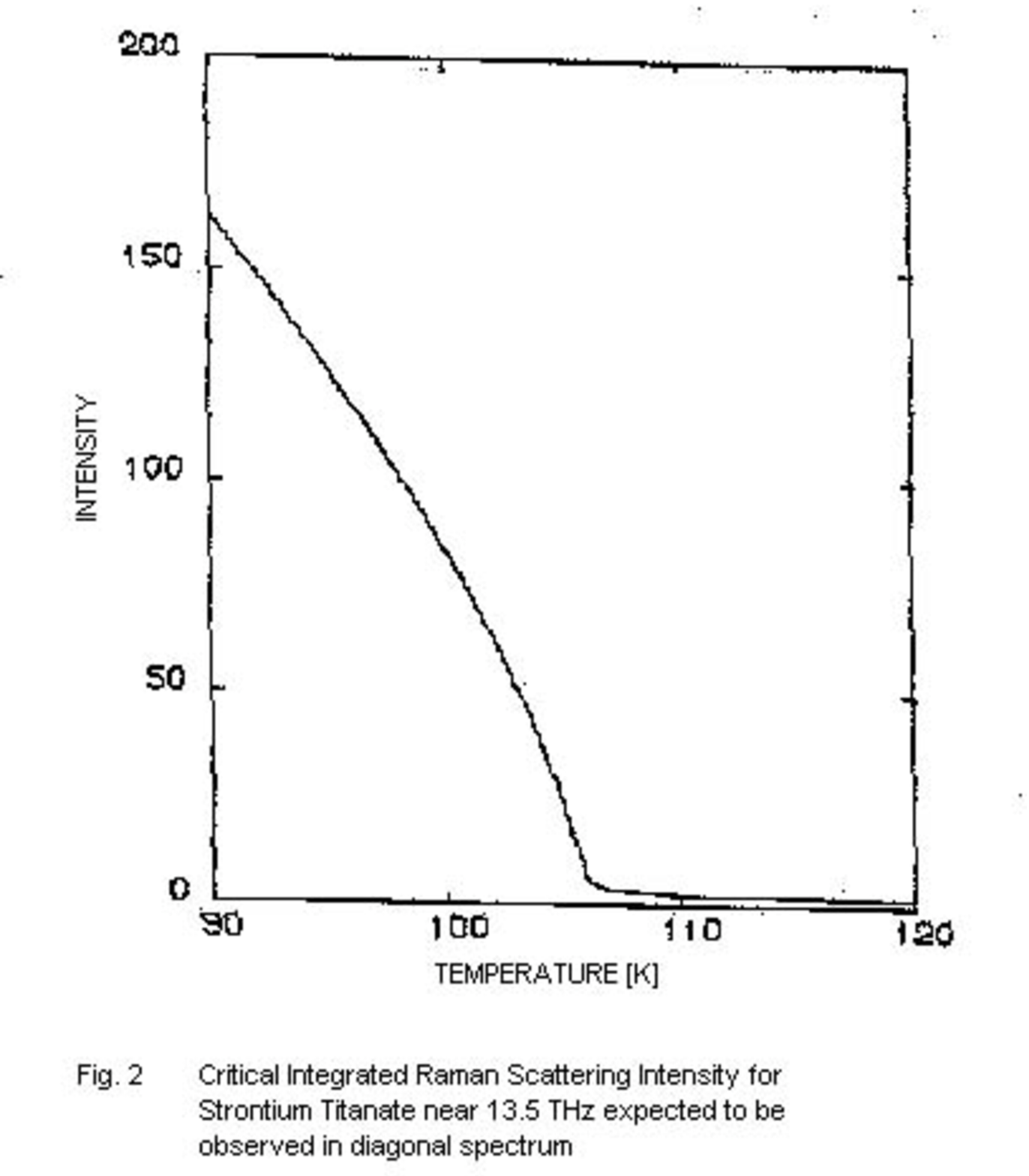, width=8.0cm, height=8.0cm}}
{\epsfig{file=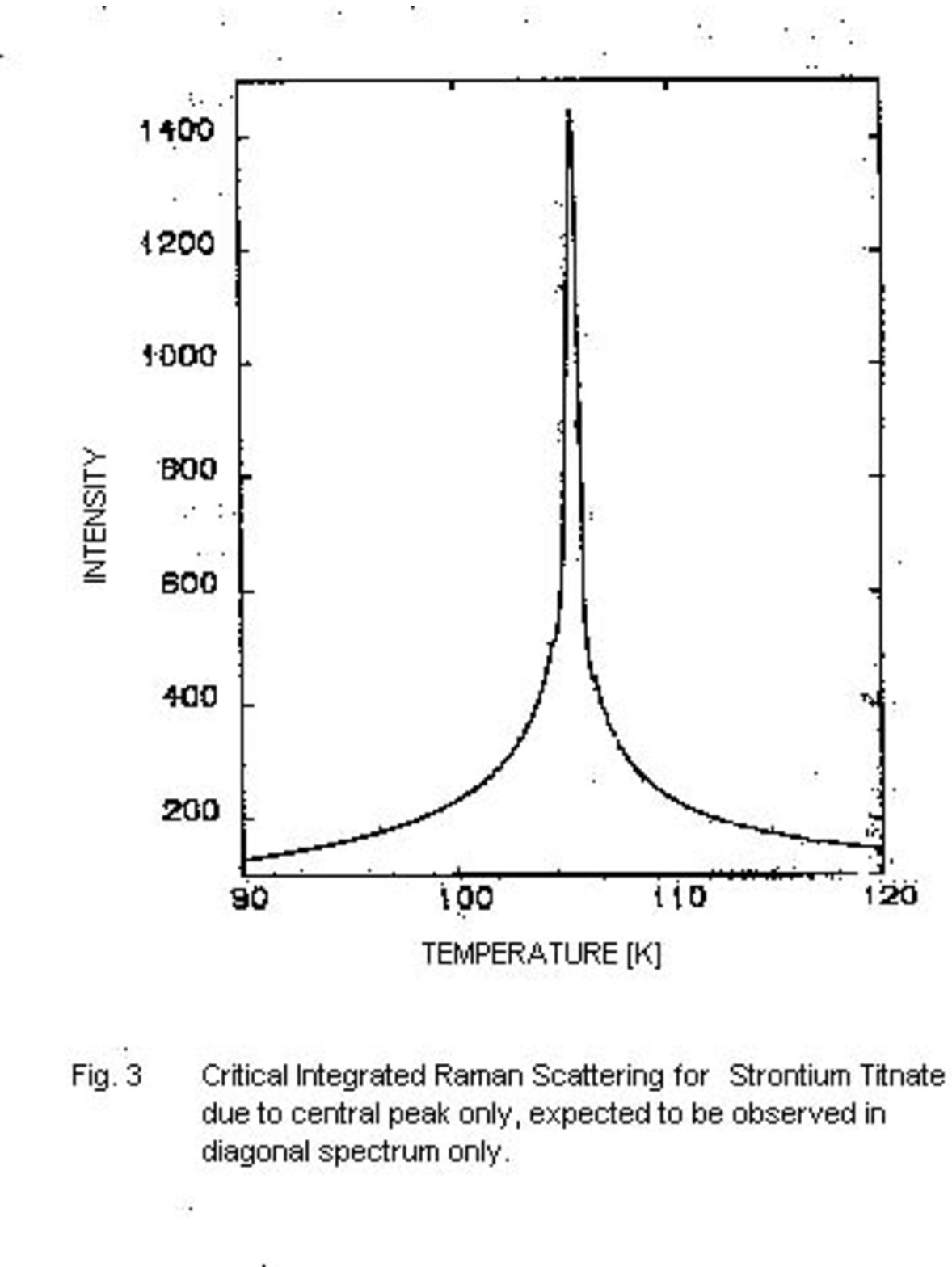, width=8.0cm, height=8.0cm}}
\end{center}
\end{figure}


\newpage
\begin{thebibliography}{99}

\bibitem{Anderson71} A. Anderson,  (Ed.) (1971) The Raman Effect, Vol. I \& II
                        (Marcel Dekker, Inc., New York).
\bibitem{Baker72} A. S. Barker  Jr and R. Loudon  (1972)
                        Rev. Mod. Phys. {\bf44} 18
\bibitem{Hayes78} W.  Hayes  and R. Loudon  (197)8 Scattering of Light by Crystals
                        (New York: Wiley)
\bibitem{Agranovich82} V. M. Agranovich  and D. L. Mills  (eds) (1982)
                        Surface Polaritons (Amsterdam: North-Holland).
                        Raman scattering by bulk and surface polaritons 4107
\bibitem{Agranovich84} V. M. Agranovich  and R. Loudon  (eds) (1984)
                        Surface Excitations (Amsterdam: North-Holland)
\bibitem{Cottam89} M.G. Cottam  and D. R. Tilley  (1989)
                        Introduction to Surface and Superlattice Excitations
                         (Cambridge: Cambridge University Press)

\bibitem{Cardona91} M. Cardona  and G. Guntherodt  (eds) (1991)
                        Light Scattering in Solids VI (Topics in
                        Applied Physics 68) (Berlin:Springer)
\bibitem{Nkoma99}  J. S. Nkoma  (1999) J. Phys.: Condens. Matter {\bf 11}  4093-4107
                        and reference cited therein.  
\bibitem{Born56} M. Born  and  K. Huang (1956) Dynamical Theory of Crystal
                Lattices, Oxford.
\bibitem{Sharma81} Jyoti Dhar Sharma, (1981) M. Phil. Thesis, University 
                        of Edinburgh.
\bibitem{Loudon64} R. Loudon {1964} Adv. Phys. {\bf 13}, 423.
\bibitem{Cowley64} R. A.Cowley  (1964) Proc. Phys. Soc., London, {\bf 84}, 281.
\bibitem{BrucePhD} A. D. Bruce  (1973) Ph. D. Thesis, University of Edinburgh.
\bibitem{Loudon63} R. Loudon  (1963) Proc. Roy. Soc., London, {\bf A275}, 218.
\bibitem{Born47}   M. Born  and  B. Bradburn (1947) Proc. Roy. Soc. {\bf A183}, 161.
\bibitem{Dick58}  B. J. Dick,  and A. W. Overhauser,  (1958) Phys. Rev. {\bf 112}, 90.
\bibitem{Cochran60} W. Cochran  (1960) Proc. Roy. Soc. {\bf A275}, 218.
\bibitem{Woods60} A. D. B. Woods  et. al. (1960) Phys. Rev. {\bf 119}, 980.
\bibitem{Bruce71} A. D. Bruce  and R. A. Cowley  (1971)
                  Indian J. Pure and Appl. Phys. {\bf9}, 877.
\bibitem{Bruce72} A. D. Bruce  and R. A. Cowley  (1972) 
                      J. Phys. C: Solid State Phys. {\bf5}, 595.
\bibitem{Schroder66} U. Schr$\ddot{o}$der  (1966), Solid St. Commun. {\bf4}, 347.
\bibitem{Nusslein67} V. Nusslein  and U. Schr$\ddot{o}$der  (1967)
                        Phys. Status Solidi, {\bf21}, 309.
\bibitem{Stirling72} W. G. Stirling  (1972)
                        J. Phys. C: Solid State Phys. {\bf5}, 2711.
\bibitem{Migoni76} R. Migoni , K. H. Rieder,  K. Fischer  and H. Bilz, 
                        (1976) Ferroelectrics {\bf13}, 377.
\bibitem{Bruce80}       A. D. Bruce,  W. Taylor  and A. F. Murray  (1980)
                         J. Phys. C: Solid State Phys. {\bf13}, 483.
\bibitem{Schwabl72}  F. Shawabl (1972) Z. Phys. {\bf254}, 57.
\bibitem{Schwabl74}  F. Schwabl  in T. Riste  (Ed.) (1974) Anharmonic Lattices,
                        Structural Phase Transitions and Melting, Noordhoff.

\bibitem{Abrikosov63} A. Abrikosov, L. P. Gorkov  and I. Ye Dzyaloshinskii, 
                        (1963) Methods of Quantum Field Theory in Statistical
                        Physics(Prentice Hall Inc.).
\bibitem{Maradudin62} A. A. Maradudin and A. E. Fein (1962) Phys. Rev. {\bf 128}, 2589.
\bibitem{Cowley63} R. A. Cowley  (1963) Adv. Phys. {\bf 12}, 421.
\bibitem{Cowley66} R. A. Cowley  in R. W. H. Stevenson  (Ed) 1966,
                        Phonons in Perfect
                        Lattices with Point Imperfections (Oliver \& Boyd).
\bibitem{Coombs73} G. J. Coombs  and R. A. Cowley  (1973)
                        J. Phys. C: Solid St. Phys. C: Solid State Phys. {\bf 6}, 121.
\bibitem{Tani69}    K. Tani  (1969)J. Phys. Soc. Japan Suppl. {\bf 26}, 93.
 \bibitem{Cowley70}    R. A. Cowley  (1970)J. Phys. Soc. Japan Suppl. {\bf 28}, S239.
\bibitem{Mori65}    H. Mori  (1965) Prog. Theor. Phys. (Kyoto) {\bf 33}, 423.
\bibitem{Cowley64a} R. A. Cowley  (1964) Phys. Rev. 134, A981.
\bibitem{Cowley69} R. A. Cowley , W. J. L. Buyers  and G. Dolling  (1969)
                   Solid State Commun, {\bf 7}, 181.
\bibitem{StirlingPhD} W. G. Stirling  (1972), Ph. D. Thesis, University of
                  Edinburgh.
\bibitem{Shapiro72} S. M. Shapiro,  J. D. Axe, G. Shirane  and T. Riste 
                         (1972) Phys. Rev. {\bf B6}, 4332.
\bibitem{Neilsen68} W. G. Neilsen  and J. G. Skinner,  (1968) J. Chem. Phys.
                {\bf 48}, 2240.
\bibitem{Fleury68} P. A. Fleury,  and J. M. Worlock,  (1968) Phys. Rev. {\bf 174}, 613.
\bibitem{Taylor79} W. Taylor,  and A. F. Murray,  (1979)
                        Solid State Commun. {\bf 31},   937.
\bibitem{Muller71} K. A. M$\ddot{u}$ller  and W. Berlinger,  (1971)
                        Phys. Rev. Lett. {\bf 26}, 13.
\bibitem{Darlington75} C. N. W. Darlington, W. J. Fitzgerald  and
                        D. A. O'Connor, 1975) Phys. Lett. {\bf 21}, 16.
\bibitem{Darlington76}C. N. W. Darlington,  and D. A. O'Connor, 
                         J. Phys. C: Solid State Phys. {\bf 9}, 3561.
\bibitem{Unoki67} H. Unoki  and T. Sakudo  (1967) J. Phys. Soc. Japan {\bf 23}, 546.
\bibitem{Fleury68a}  P. A. Fleury, J. F. Scott,  and J. M. Worlock,  (1968)
                 Phys. Rev. Lett. {\bf 21}, 16.
\bibitem{Fleury67} P. A. Fleury and  J. M. Worlock  (1967)
                        Phys. Rev. Lett. {\bf 18}, 665.
\bibitem{Worlock67}  J. M. Worlock  and  P. A. Fleury (1967) Phys. Rev. {\bf 19}, 1176.
\bibitem{Shirane69} G. Shirane  and  T. Yamada (1969) Phys. Rev. {\bf 117}, 858.
\bibitem{Yacoby76} Y. Yacoby,  W. W. Kruhler  and S. Just  (1976)
                Phys. Rev. {\bf B13}, 4132.
\bibitem{Riste71} T. Riste,  E. J. Samuelsen,  R. Otnes,  and  J. Feder,
                (1971) Solid. State Commun. {\bf 9}, 1455.
\bibitem{Topler77} J. T$\ddot{o}$pler,  B. Alefeld,  and A. Heidemann  (1977)
                J. Phys. C: Solid State Phys. {\bf 10}, 635.
\bibitem{Cowley78} R. A. Cowley,   and G. Shirane,  (1978) J. Phys. C: Solid
                        State Phys. {\bf 11}, L939.
\bibitem{Nara61}  P. S. Narayanan,  and K. Vedam, (1961) Z. Phys. {\bf 163}, 158.
\bibitem{Heine60}  V. Heine,  (1960) Group Theory in Quantum Mechanics
                (Pergamon, New York).
\bibitem{Tinkham64} M. Tinkham (1964) Group Theory an Quantum Mechanics,
                (McGraw-Hill, New York).
\bibitem{Cowley71a} R. A. Cowley  in  A. Anderson (Ed.) (1971)
                The Raman Effect, Vol. I (Marcel Dekker, Inc., New York).
\bibitem{Lockwood74} D. J. Lockwood  and B. H. Torrie  (1974)
                 J. Phys. C: Solid State Phys. {\bf 7}, 2729
\bibitem{Koster57} G. F. Koster  (1957) Solid St. Phys. {\bf 5}, 174.
\bibitem{Riste71a} T. Riste, E. J. Samuelson and K. Otnes (1971) Structural
Phase Transitions and Soft Modes, Edited by E. J. Samuelson, E. Anderson and
J. Feder. (Oslo Universitetsforlaget).pp.395-408.
\bibitem{Cowley80a} R. A. Cowley(1980)Adv. Phys. {\bf 29}, 1-110. p.19
\bibitem{Courtens72} E. Courtens (1972) Phys. Rev. Lett. {\bf 29}, 1380.
\bibitem{Bruce80a} A. D. Bruce (1980)Adv. Phys. {\bf 29}, 111-217. p.195




\end{thebibliography}
\end{document}